\documentclass[aps,prl,twocolumn,superscriptaddress,longbibliography]{revtex4-2}
\usepackage[T1]{fontenc}
\usepackage[utf8]{inputenc}
\usepackage{amsmath,amssymb,mathtools,bm}
\usepackage{graphicx}
\usepackage{xcolor}
\usepackage{hyperref}
\usepackage{microtype}
\usepackage{times}

\newcommand{\pc}{p_{\mathrm c}}
\newcommand{\popen}{p_{\mathrm c}^{\mathrm{open}}}

\begin{document}

\title{Disorder Crossover in Urban-Front Growth}
% Universal Local Roughness from Disorder Crossover in Urban-Front Growth}
% Disorder Crossover Explains Urban-Front Roughening

\author{Martin Hendrick}
\affiliation{Laboratory of Urban and Environmental Systems, \'{E}cole Polytechnique F\'{e}d\'{e}rale de Lausanne,\\ Lausanne, Switzerland}
\author{Maximilian Trique}
\affiliation{Laboratory of Urban and Environmental Systems, \'{E}cole Polytechnique F\'{e}d\'{e}rale de Lausanne,\\ Lausanne, Switzerland}
% \author{...}
\author{Gabriele Manoli}
\affiliation{Laboratory of Urban and Environmental Systems, \'{E}cole Polytechnique F\'{e}d\'{e}rale de Lausanne,\\ Lausanne, Switzerland}

\date{\today}

%\begin{abstract}
%Recent measurements of urban expansion fronts revealed a robust local roughness exponent together with strongly dispersed growth and dynamic exponents. We show that this combination arises from a disorder-controlled crossover in projected-front growth. To capture the empirical ingredients at minimal cost, we introduce an Eden model in which geographic constraints act as quenched dilution, while coalescence and preferential development act as quenched local acceleration. Because the disorder is isotropic and blocked regions can be bypassed by the growing cluster, the near-threshold geometry is governed by ordinary two-dimensional percolation rather than directed-percolation depinning. The projected front therefore spends long times in a disorder-dominated preasymptotic regime before crossing over to the asymptotic KPZ sector. In this regime, the local roughness remains close to $1/2$, whereas the large-scale dynamics produces broad apparent growth and dynamic exponents. This mechanism provides a minimal explanation of urban-front roughening and may apply more generally to stochastic growth in heterogeneous media.

%\end{abstract}

\begin{abstract}
Urban expansion fronts display a robust local roughness exponent together with strongly dispersed growth and nonuniversal dynamic exponents. We show that this coexistence can arise from a disorder-controlled crossover in projected-front growth. Introducing a minimal Eden model, in which geographic constraints act as quenched dilution and coalescence as quenched local acceleration, we demonstrate that the resulting front enters a long disorder-dominated preasymptotic regime, whose scaling near threshold is set by ordinary two-dimensional percolation. In this regime, the local roughness remains close to $1/2$, while the large-scale exponents vary broadly with disorder and acceleration. These results provide a minimal explanation of urban-front roughening and suggest a more general mechanism for stochastic growth in heterogeneous media.
\end{abstract}

\maketitle

Advancing rough fronts occur across a wide range of nonequilibrium systems, from biological colony growth~\cite{huergo2010morphology,swartz2024new} and range expansions~\cite{chu2019evolution,horowitz2019bacterial} to electrodeposition~\cite{schilardi1999validity,kahanda1992columnar}, thin-film deposition~\cite{almeida2014universal,orrillo2017morphological}, and combustion~\cite{myllys2001kinetic}. A central problem in all these settings is to determine whether measured exponents reflect a genuine asymptotic universality class or instead a long crossover generated by quenched heterogeneity, branching, coalescence, or competing transport channels. A standard starting point for this question is provided by local stochastic growth models, including Eden-type rules, and by their coarse-grained KPZ description~\cite{eden1961two,kardar1986dynamic}. Yet several important examples already show that finite-time observations can be dominated by extended preasymptotic sectors: disorder-dependent effective exponents in KPZ-related growth may arise from slow crossover toward universal asymptotic values rather than genuine nonuniversality~\cite{de1999universality}; electrodeposition models display unstable or anomalous roughening before crossing over to asymptotic KPZ scaling~\cite{castro1998anomalous,castro2000multiparticle}; environmental heterogeneity can qualitatively alter front-mediated evolutionary dynamics in range expansions~\cite{gralka2019environmental}; and mutation-selection couplings in two-species Eden growth feed back on interface roughening and produce exponents distinct from those of the clean Eden model~\cite{kuhr2011range}.

Urban expansion represents a novel and fertile playground for studying the dynamics of nonequilibrium systems~\cite{marquis2025universal}. While an important problem in its own right, as humanity now predominantly lives in cities~\cite{worldbank_urban,unhabitat_wcr2022,un_urbanization2025}, urbanisation has also become a central object of study for statistical physics, which has revealed robust regularities in urban systems and motivated dynamical, coarse-grained descriptions of their behaviour~\cite{barthelemy2019statistical,bettencourt2019towards}. Recent nonequilibrium approaches have further emphasized urban growth as an emergent stochastic process with collective scaling properties~\cite{bettencourt2020urban}. Urban sprawl therefore offers an unusually clean empirical setting in which to confront the broader problem of rough-front dynamics in heterogeneous media.

A recent study analyzed the dynamics of the largest connected urban component in 19 metropolises between 1985 and 2015, using the World Settlement Footprint Evolution dataset to reconstruct the yearly growth of built-up areas~\cite{marquis2025universal,wsf_evolution}. Based on radial roughness observables of these evolving fronts, this work found two striking facts: the local roughness exponent is nearly constant, $\alpha_{\rm loc}\simeq0.54$, whereas the growth exponent $\beta$ and the dynamic exponent $z$ vary strongly from city to city, with no city fitting clearly into a standard clean or quenched interface universality class~\cite{marquis2025universal}. The same study further showed that demographic pressure changes the dominant expansion mechanism: low pressure favours local agglomeration, whereas high pressure enhances coalescence with surrounding built clusters~\cite{marquis2025universal}. The theoretical question is therefore sharp: what minimal growth framework can produce a robust short-scale roughness exponent together with broad, nonuniversal effective values of $\beta$ and $z$?

Here we argue that a minimal explanation is provided by a projected-front Eden model with two quenched ingredients: dilution, which mimics geographic accessibility constraints, and local acceleration, which mimics coalescence and preferential development corridors. The key conceptual point is that the exponents reported in Ref.~\cite{marquis2025universal} need not be interpreted as belonging to different asymptotic universality classes. Instead, they can arise from a single disorder-controlled crossover regime preceding the eventual moving-phase KPZ sector. Near the dilution threshold the front morphology undergoes the isotropic channelization transition identified by Grassberger for critically pinned interfaces in random media~\cite{grassberger2018universality}. Within this picture, dilution fixes the accessibility geometry, acceleration modifies the traversal-time metric on that geometry and imprints an early correlated front, and the observed coexistence of nearly universal $\alpha_{\rm loc}$ with variable $\beta$ and $z$ becomes natural.

Although we formulate the theory in Eden language because it is the minimal local stochastic growth rule, the underlying mechanism may be more general. It requires an overhang-rich growth morphology, a projected single-valued front observable, and a quenched distinction between accessibility disorder and traversal-time disorder. Under local dynamics with short-range fluctuations, these ingredients naturally produce Brownian-like local roughness together with broad effective dynamical exponents generated by disorder crossover. This suggests that related behavior may occur in other stochastic growth problems, including some colony expansions, range expansions, and deposition processes, although systems with long-range interactions, correlated disorder, or genuinely nonlocal growth rules need not follow the same scenario.

\begin{figure}[t]
\centering
\includegraphics[width=7.5cm]{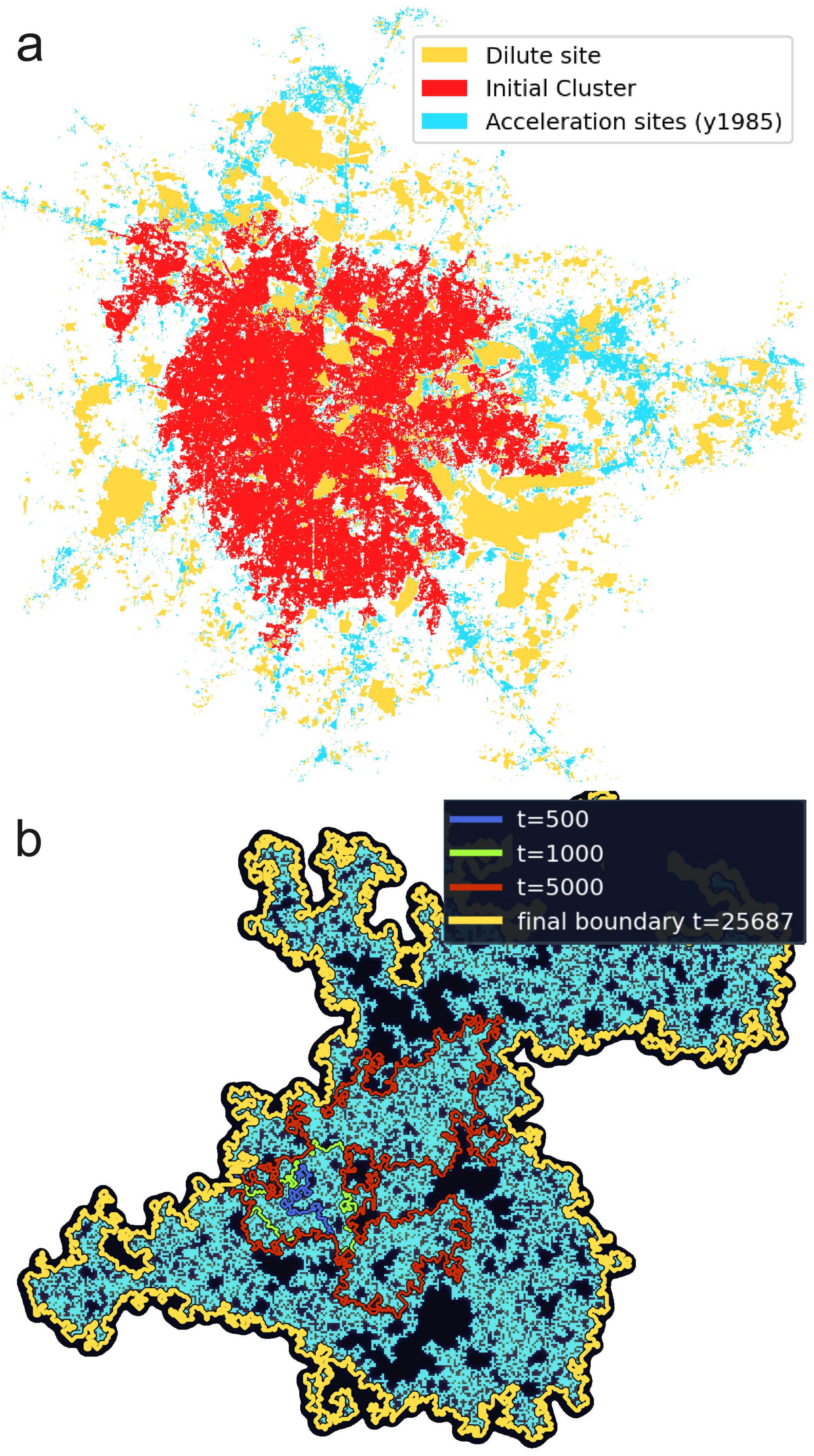}
\caption{(a) Largest connected urbanized area of Bangalore (India) in 1985 (red). Yellow sites indicate areas that remain nonurbanized in 2015 and are identified with diluted sites, while blue sites indicate urbanized patches that are disconnected from the main cluster in 1985 but become absorbed into it by 2015; these are identified with acceleration sites. (b) Representative morphologies of the diluted--accelerated Eden model with quenched dilution and quenched acceleration. Black regions denote unoccupied sites, and colored lines show the cluster boundary at successive times.}
\label{fig:morph}
\end{figure}

We consider Eden growth~\cite{eden1961two} on a square lattice with blocked sites of probability $p$ and accelerated sites of conditional probability $q$ among the open sites. The correspondence with urban data is illustrated in Fig.~\ref{fig:morph}(a), while Fig.~\ref{fig:morph}(b) shows a representative morphology generated by the diluted--accelerated Eden model (see SM for details about model's implementation and related observable computation and Figs.\ref{fig_si:illu_radial} and Figs.\ref{fig_si:illu_strip} for some representative simulations). The front analyzed in practice is a projected, single-valued observable. In strip geometry one may define
\begin{equation}
 h(x,t)=\max\{y:(x,y)\ \text{occupied at time}\ t\},
\label{eq:hx}
\end{equation}
while in radial geometry, more directly relevant to cities, one uses
\begin{equation}
 R(\theta,t)=\max\{r:(r,\theta)\ \text{occupied at time}\ t\}.
\label{eq:rtheta}
\end{equation}
Local widths are then computed in windows of arc length $\ell\simeq \bar R(t)\Delta\theta$, in the same spirit as in Ref.~\cite{marquis2025universal}. All exponents discussed below belong to this projected-front observable, not to the full multiply valued outer hull. In this work we analyze the front mainly in strip geometry, which offers a fixed and unambiguous system size and therefore allows a cleaner determination of correlation lengths, crossover scales, and finite-size cutoffs than radial geometry.

\emph{Clock choice}.—The choice of clock is essential. In the urban data, the natural monotone control variable is the population $P$ of the largest connected component rather than calendar time~\cite{batty2008size,marquis2025universal}, estimated through $P\simeq A\,P_{\rm tot}/A_{\rm tot}$~\cite{marquis2025universal}. To remain as close as possible to this procedure, we use as our primary clock the cluster mass $t_{\rm mass}\equiv M$, the natural model analogue of population. We also analyze the same dynamics in terms of the continuous-time clock
\begin{equation}
t_{\rm ct}(n)=\sum_{m=1}^{n}\frac{1}{N_{\rm perim}(m)},
\label{eq:clock}
\end{equation}
where $N_{\rm perim}(m)$ is the number of active perimeter sites before the $m$th stochastic event. This accumulated mean waiting time is the natural physical time of the asynchronous Eden process and, near critical dilution, is especially useful because it probes local propagation along the chemical paths of the critical open cluster. Unless otherwise stated, we write simply $t$ and specify explicitly whether $t=t_{\rm mass}$ or $t=t_{\rm ct}$ whenever the clock choice matters.

\emph{Dilution and acceleration}.—The first quenched field, dilution, controls accessibility. For random site dilution, the associated critical geometry is that of ordinary two-dimensional site percolation~\cite{stauffer2018introduction}. Writing $\Delta=\pc-p>0$, with $\pc=1-\popen\approx0.407254$ on the square lattice, the disorder length obeys
\begin{equation}
 \xi_p\sim \Delta^{-\nu},\qquad \nu=\frac{4}{3}.
\label{eq:xip}
\end{equation}
The second quenched field, acceleration, modifies motion on the already accessible open cluster. If accelerated sites are sampled independently among open sites, their usable density is $\rho_a=(1-p)q$. The accelerated subnetwork percolates when
\begin{equation}
 (1-p)q=\popen,
\qquad q_{\mathrm{acc}}(p)=\frac{\popen}{1-p}.
\label{eq:qacc}
\end{equation}
For $q<q_{\mathrm{acc}}(p)$, accelerated clusters are finite and leave accessibility unchanged, but they alter the traversal-time metric and imprint short-scale correlations on the projected front. For $q>q_{\mathrm{acc}}(p)$, touching a spanning accelerated cluster triggers effectively instantaneous avalanches, so conventional exponent extraction becomes clock dependent and the notion of a smooth roughening regime may break down.

\begin{figure*}[t]
\centering
\includegraphics[width=1\textwidth]{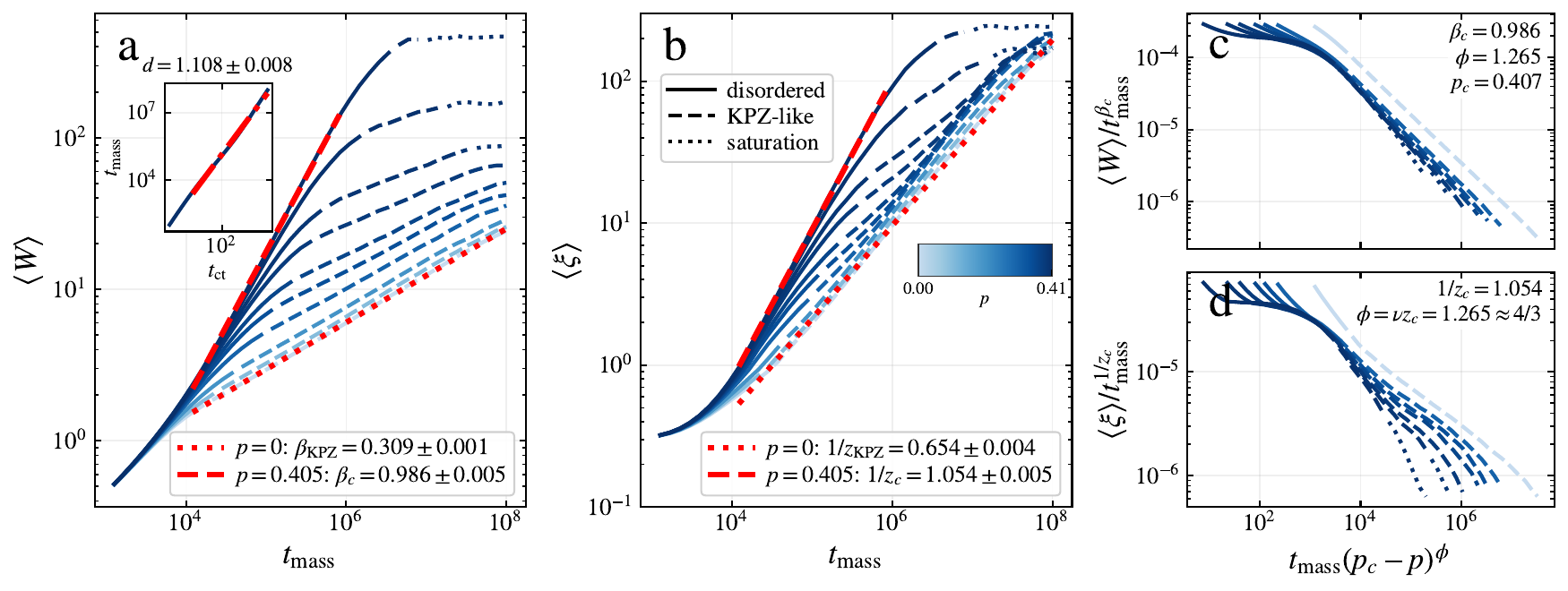}
\caption{Global-width and correlation-length scaling in the dilution-only case $q=0$. Data are averaged over $500$ realizations for a strip of width $L=5000$. (a) Global width $W(t_{\rm mass})$ for representative $p<p_c$. The clean case $p=0$ is consistent with KPZ growth, while increasing dilution extends the disorder-dominated transient. Inset: clock conversion $t_{\rm mass}\sim t_{\rm ct}^{1/d}$, with fitted exponent $d\simeq1.108\approx d_{\min}$. (b) Correlation length $\xi(t_{\rm mass})$ for the same values of $p$. (c) and (d) Collapses of the width data according to, respectively, Eq.~\eqref{eq:Wscaling} and Eq.~\eqref{eq:xiscaling} for $p\ge 0.3$, $p=0$ is shown for reference.}
\label{fig:width}
\end{figure*}

\emph{Disorder-dominated regime}.—We first focus on the subcritical regime $p<p_c$ and $q<q_{\mathrm{acc}}(p)$, where accessibility is controlled by the open cluster and acceleration acts only through finite fast regions. The projected global width is then expected to display a disorder-dominated transient regime,
\begin{equation}
W(t)\sim t^{\beta_d},
\qquad
t\ll t_{\mathrm{cross}},
\label{eq:Wdis}
\end{equation}
followed, when the disorder scale is sufficiently smaller than the maximal dynamically generated correlation length, by a KPZ-like regime
\begin{equation}
W(t)\sim t^{1/3},
\qquad
t_{\mathrm{cross}}\ll t\ll t_{\mathrm{fs}}.
\label{eq:Wkpz}
\end{equation}
If this scale separation is not realized, the KPZ window may be bypassed altogether and the growth crosses directly to saturation or finite-size cutoff. Here $t_{\mathrm{cross}}$ marks the end of the disorder-dominated transient, $t_{\mathrm{fs}}$ is a later geometry- or size-dependent cutoff, and $\beta_d$ is an effective transient exponent. These different regimes are shown from simulation results in Fig.~\ref{fig:width}(a).

\emph{Dilution-only crossover}.—We first consider the dilution-only case, $q=0$, in order to isolate the crossover generated by accessibility disorder alone. To characterize the disorder-dominated regime, we extract a dynamic correlation length from the height-height correlation function, Fig.~\ref{fig:width}(b). In the transient window,
\begin{equation}
 \xi(t)\sim t^{1/z_d},
 \qquad
  t\ll t_{\mathrm{cross}},
\label{eq:xit}
\end{equation}
where $z_d$ is an effective dynamical exponent. The crossover out of this regime occurs when the front correlation length reaches the percolative disorder scale,
\begin{equation}
 \xi(t_{\mathrm{cross}})\sim \xi_p.
\label{eq:match}
\end{equation}
Using $\xi_p\sim \Delta^{-\nu}$, one obtains
\begin{equation}
 t_{\mathrm{cross}}\sim \xi_p^{z_d}\sim \Delta^{-\phi},
 \qquad
 \phi=\nu z_d.
\label{eq:phi_general}
\end{equation}
A convenient scaling ansatz for the projected global width is then
\begin{equation}
 W(t,\Delta)\sim t^{\beta_c}\,\mathcal F\!\left(t\Delta^{\phi}\right),
\label{eq:Wscaling}
\end{equation}
while the correlation length obeys the companion form
\begin{equation}
\xi(t,\Delta)\sim t^{1/z_c}\,\mathcal X\!\left(t\Delta^{\phi}\right).
\label{eq:xiscaling}
\end{equation}
For $u=t\Delta^\phi\ll1$, the system remains in the disorder-controlled regime; for $u\gg1$, it crosses over toward the moving phase. Thus different effective distances from the morphological transition naturally generate broad drifts in the apparent exponents. Figure~\ref{fig:width}(c,d) confirms both scaling forms in the mass clock.

\emph{Critical scaling hypothesis}.—Near the dilution threshold, it is natural to expect that the spreading of correlations in the projected front is controlled by chemical transport on the critical open cluster. If two front points are separated by a Euclidean distance $r$, the shortest accessible path scales as $\ell_{\min}(r)\sim r^{d_{\min}}$, where $d_{\min}=1.13077(2)$ is the ordinary-percolation shortest-path exponent~\cite{stauffer2018introduction}. Measured in the propagation clock $t_{\rm ct}$, this suggests $t_{\rm ct}\sim \ell_{\min}\sim r^{d_{\min}}$, and therefore
\begin{equation}
\xi(t_{\rm ct})\sim t_{\rm ct}^{1/d_{\min}},
\qquad
z_c^{(\rm ct)}\simeq d_{\min},
\qquad
\phi=\nu d_{\min}\approx1.51.
\label{eq:zc_phi}
\end{equation}
The same percolative picture, together with Grassberger's line-seed result for isotropically pinned interfaces~\cite{grassberger2018universality}, suggests that the mean projected height obeys
\begin{equation}
\bar h(t_{\rm ct})\sim t_{\rm ct}^{1/d_{\min}}.
\label{eq:hbar}
\end{equation}
Under the single-scale ansatz
\begin{equation}
P(h,t_{\rm ct})=t_{\rm ct}^{-1/d_{\min}}f\!\left(\frac{h}{t_{\rm ct}^{1/d_{\min}}}\right),
\label{eq:single}
\end{equation}
one then obtains
\begin{equation}
W(t_{\rm ct})\sim t_{\rm ct}^{\beta_c^{(\rm ct)}},
\qquad
\beta_c^{(\rm ct)}=\frac{1}{d_{\min}}\approx0.884,
\label{eq:betac}
\end{equation}
see Fig.~\ref{fig_si:pdf_h_collapse} in SM for a numerical validation. Independently, the inset of Fig.~\ref{fig:width}(a) shows that in our simulations the mass clock is related empirically to the propagation clock by
\begin{equation}
t_{\rm mass}\sim t_{\rm ct}^{1/d},
\qquad d\simeq1.108\approx d_{\min}.
\label{eq:tmass_tct}
\end{equation}
Rewriting the same critical laws in terms of $t_{\rm mass}$ then gives
\begin{equation}
W(t_{\rm mass})\sim t_{\rm mass},
\qquad
\xi(t_{\rm mass})\sim t_{\rm mass},
\label{eq:mass_scaling}
\end{equation}
so that
\begin{equation}
\beta_c^{(\rm mass)}\simeq1,
\qquad
\frac{1}{z_c^{(\rm mass)}}\simeq1.
\label{eq:mass_exp}
\end{equation}
Thus the numerical values of $\beta_c$ and $z_c$ depend on the chosen clock once quenched disorder is present: $t_{\rm ct}$ exposes the chemical geometry of critical transport, whereas $t_{\rm mass}$ is the natural parametrization for comparison with urban fronts. Beyond the critical-point analysis, Fig.~\ref{fig_si:beta_invz_vs_p} in the Supplemental Material shows that the effective exponents measured in the disorder-dominated regime for $q=0$ vary continuously with $p$. Here the fit window is restricted to the post-Poissonian and pre-KPZ/pre-saturation sector, namely before $d\ln W/d\ln t$ drops below $1/3$ for $\beta_d$ and before $d\ln \xi/d\ln t$ drops below $2/3$ for $1/z_d$. This continuous drift is consistent with the crossover interpretation developed here: away from $p_c$, the projected front is governed by effective disorder-controlled exponents rather than by a sharp asymptotic critical law.

\begin{figure*}[t]
\centering
\includegraphics[width=1\textwidth]{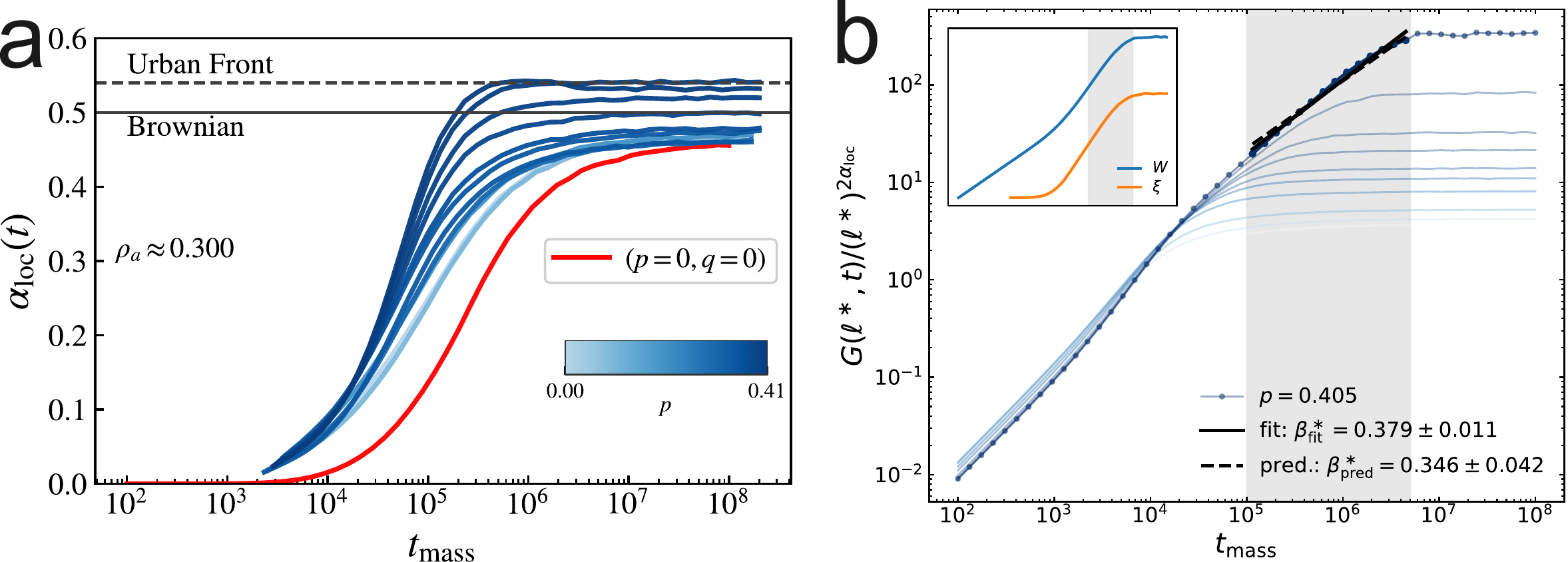}
\caption{Local-sector dynamics of the projected front. (a) Time evolution of $\alpha_{\rm loc}$ in the mass clock. The clean case $(p,q)=(0,0)$ reaches the Brownian value only late in the evolution, close to saturation, whereas a quenched fields with e.g. $\rho_a\approx0.3$ drives $\alpha_{\rm loc}$ rapidly toward $1/2$ for representative dilution strengths $p<p_c$. (b) Compensated local correlation in the dilution-only case $q=0$ for several values of $p<p_c$, colored by $p$. The gray band marks the fitting window; the emphasized curve corresponds to the largest $p$. Solid and dashed black lines show the fitted and predicted local-growth laws, respectively. Inset: normalized $W(t)$ and $\xi(t)$ for the same largest-$p$ dataset.}
\label{fig:local}
\end{figure*}

\emph{Robust local roughness}.—Why then does the local exponent remain so close to $1/2$ despite these strong changes in the large-scale dynamics? The reason is that local roughness is controlled by the short-distance sector of the projected front, whereas dilution and acceleration affect most strongly the long-wavelength dynamics. In the generic anomalous-scaling picture, the height-height correlation and the local width obey
\begin{equation}
G(\ell,t)=\langle [h(x+\ell,t)-h(x,t)]^2\rangle\sim \ell^{2\alpha_{\rm loc}},
\qquad
w(\ell,t)\sim \ell^{\alpha_{\rm loc}},
\label{eq:Gdef}
\end{equation}
at small $\ell$, where $w$ denotes the local width~\cite{ramasco2000generic}. In our simulations, the disorder fields strongly modify the infrared sector while the measured short-distance behavior remains consistent with local increments $\delta h_x=h(x+1,t)-h(x,t)$ that have finite variance and only short-range correlations. In that case
\[
h(x+\ell,t)-h(x,t)=\sum_{j=0}^{\ell-1}\delta h_{x+j}
\]
has a variance that grows approximately linearly with $\ell$, implying
\begin{equation}
G(\ell,t)\sim \ell,
\qquad
w(\ell,t)\sim \ell^{1/2}.
\label{eq:brownian}
\end{equation}
In the clean case, the projected front starts from a comparatively smooth profile, and the Brownian short-distance sector develops only gradually. As a result, the scaling window $\ell\ll \xi(t)$ remains narrow for a long time and the measured $\alpha_{\rm loc}$ shows strong preasymptotic drift, approaching $1/2$ only when the interface is close to its stationary spatial statistics, which in finite systems occurs late in the evolution, often near saturation. Moderate dilution at $q=0$ produces little visible change, whereas dilution close to $p_c$ stabilizes $\alpha_{\rm loc}\simeq1/2$ noticeably earlier. The clearest effect comes from acceleration: for $q>0$, the projected front approaches $\alpha_{\rm loc}\simeq1/2$ already at early growth stages, well before saturation, as shown in Fig.~\ref{fig:local}.

\begin{figure*}[t]
\centering
\includegraphics[width=1\textwidth]{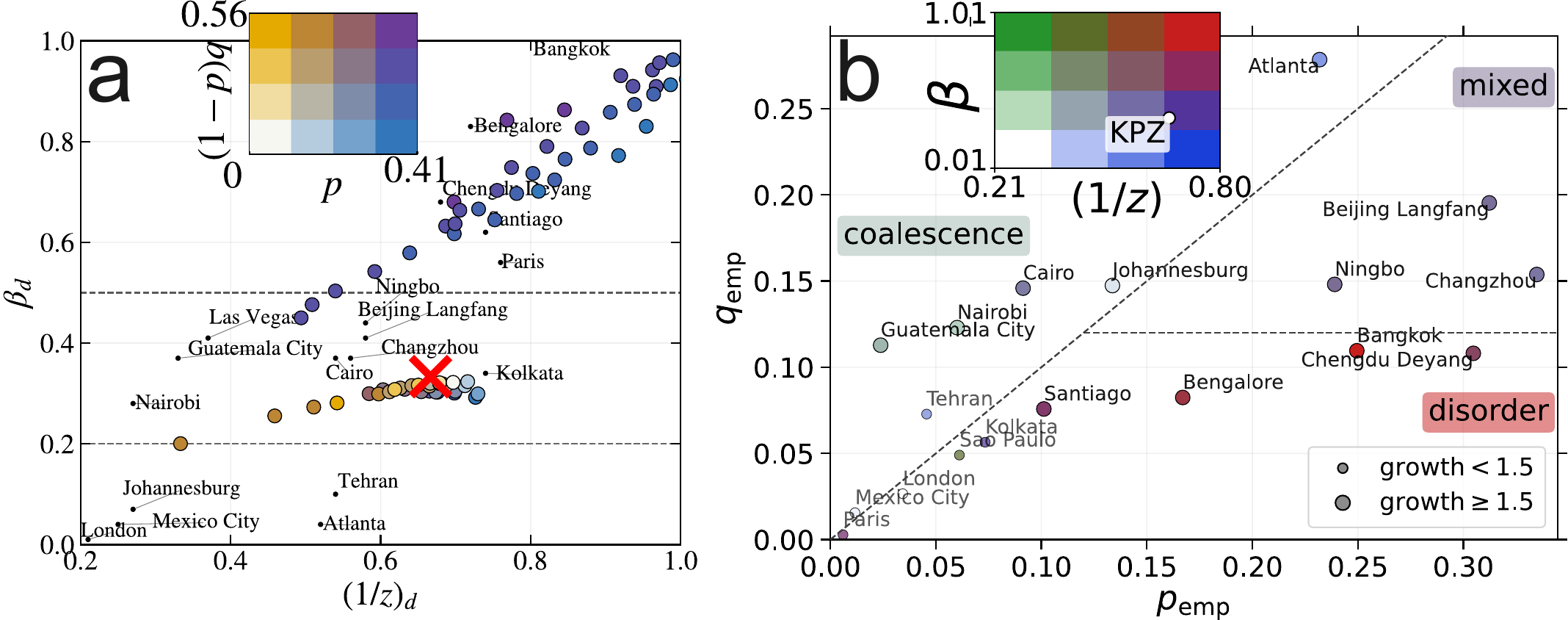}
\caption{(a) Effective $(\beta_d,1/z_d)$ values measured in the disorder-dominated regime (pre-KPZ or saturation regime) of the diluted--accelerated Eden model (colored dots) for parameters $(p,q)$ below the thresholds $p_c$ and $q_{\mathrm{acc}}(p)$, compared with representative cities from Ref.~\cite{marquis2025universal} (black points). The bivariate colormap encodes the corresponding model parameters $(p,q)$ for each simulated point. The model generates a broad cloud of apparent dynamical exponents while preserving a common local-roughness sector. The KPZ reference point is marked by the red cross. (b) Empirical proxy phase diagram in the $(p_{\rm emp},q_{\rm emp})$ plane for the cities reported in Ref.~\cite{marquis2025universal}. Colors encode the pair $(1/z,\beta)$ through a bivariate colormap, while marker size indicates the growth ratio. The lines $q_{\rm emp}=p_{\rm emp}$ and $q_{\rm emp}=0.12$ are shown as empirical boundaries separating coalescence-dominated, mixed, and disorder-dominated sectors.}
\label{fig:phase}
\end{figure*}

\emph{Effect of acceleration}.—Acceleration provides the second ingredient needed by the empirical study, namely a mechanism for broad city-to-city variability in the effective dynamical exponents without destroying the local exponent. Physically, accelerated clusters mimic coalescence with neighboring settlements and preferential development along favored corridors, as illustrated for Bangalore in Fig.~\ref{fig:morph}(a). In the model, increasing $q$ at fixed $p<\pc$ leaves the accessibility geometry unchanged but modifies the traversal-time metric on that geometry and enhances the initially correlated state of the front. Its clearest effect is on the correlation-length dynamics: the measured $1/z_d$ decreases as $q$ increases, meaning that $\xi(t)$ grows more slowly with the chosen clock (see Figure~\ref{fig:phase}(a) and see Fig. \ref{fig_si:width_acc} in SM). The reason is that accelerated regions promote rapid advance along selected channels, but this channelized motion does not efficiently build correlations across the full interface; instead, it tends to localize the growth and delay the spread of transverse correlations. By contrast, the growth exponent $\beta_d$ is affected more weakly, with at most a modest downward drift over the explored range of parameters. Thus acceleration broadens the cloud of apparent $(\beta,1/z)$ values mainly through its effect on correlation spreading, while simultaneously preparing the conditions under which the local sector can approach its Brownian regime earlier.

\emph{Local anomalous scaling}.—We now turn to the local roughening properties of the projected front and to the corresponding anomalous-scaling regime. The previous paragraph showed that the model maintains a robust local roughness exponent, $\alpha_{\rm loc}\simeq1/2$, over a broad range of quenched fields. To characterize how this local sector evolves in time, we use the generalized dynamic-scaling form~\cite{ramasco2000generic}
\begin{equation}
w(\ell,t)=t^{\beta_d}\,\mathcal G\!\left(\frac{\ell}{t^{1/z_d}}\right),
\label{eq:local_scaling}
\end{equation}
which, in the local regime $\ell\ll \xi(t)\sim t^{1/z_d}$, reduces to
\begin{equation}
w(\ell,t)\sim \ell^{\alpha_{\rm loc}} t^{\beta^\ast},
\qquad
\beta^\ast=\beta_d-\frac{\alpha_{\rm loc}}{z_d}.
\label{eq:local_urban}
\end{equation}
Equivalently, the height-height correlation obeys
\begin{equation}
G(\ell,t)\sim \ell^{2\alpha_{\rm loc}} t^{2\beta^\ast}.
\label{eq:Gcomp}
\end{equation}
The exponent $\beta^\ast$ therefore quantifies the time dependence of the local roughness sector and provides a direct measure of anomalous local growth. Figure~\ref{fig:local}(b) shows that the compensated local correlation develops a clear power-law growth as $p\to p_c$, signaling an anomalous local-growth regime associated with the percolative sector (see Fig.~\ref{fig_si:beta_star_acc} in the Supplemental Material for a representative $q>0$ case). This regime is genuinely critical at $p=p_c$, while for $0<\Delta\ll1$ it survives only as a long preasymptotic sector before the eventual crossover to KPZ. Farther below $p_c$, the compensated curves become much flatter, so the same mechanism appears mainly as an anomalous crossover effect rather than as a well-developed scaling regime. In this sense, the local sector of the model follows the same general phenomenology as the urban fronts reported in Ref.~\cite{marquis2025universal}: a robust $\alpha_{\rm loc}$ coexisting with a nontrivial, disorder-dependent local growth dynamics.

The previous results imply that the model naturally enters a transient anomalous roughening regime. Defining the effective global roughness exponent as
\begin{equation}
\alpha_d=\beta_d z_d,
\label{eq:alphad}
\end{equation}
one generally finds $\alpha_d\neq \alpha_{\rm loc}$ throughout the disorder-controlled crossover. The projected front is therefore not described by simple Family--Vicsek scaling in this regime; in particular, for $q>0$ the local sector is already close to $\alpha_{\rm loc}\simeq1/2$ while the global exponents remain strongly dependent on $(p,q)$.

% The urban interpretation is immediate. Geographic barriers, zoning, water bodies, and topography map onto dilution. Coalescence and corridor-like development map onto acceleration. Since the empirical fronts are measured over finite observation windows and with population rather than physical time as the control variable, the natural expectation is that cities are observed for effective clocks corresponding to $t\lesssim t_{\mathrm{cross}}$, namely inside the disorder-controlled crossover. In that regime $\alpha_{\rm loc}\simeq1/2$ can already be established because it is controlled by short-distance fluctuations, while $\beta$ and $z$ still drift with the long-wavelength disorder landscape. Different cities then need not realize different universality classes. They can occupy different positions inside the same crossover phase diagram.

\emph{Comparison with urban data}.—To confront the model with the urban data more directly, we introduce two empirical geometric proxies for its control parameters. We define $p_{\rm emp}$ as the density of emergent empty cavities inside the final largest connected component, which quantifies the effective dilution felt by the front, and $q_{\rm emp}$ as the density of already-built but not-yet-connected clusters inside the final LCC that can later be absorbed by the tracked lineage, which quantifies the effective coalescence reservoir (see SM for more details). As shown in Fig.~\ref{fig:phase}(b), cities placed in the $(p_{\rm emp},q_{\rm emp})$ plane organize in a way broadly consistent with the simulations: large $p_{\rm emp}$ and small $q_{\rm emp}$ are associated with $\beta$ and $1/z$ values above KPZ, as expected for a disorder-dominated crossover; large $q_{\rm emp}$ is associated with reduced $\beta$ and $1/z$, consistent with growth controlled by coalescence; and intermediate values of both proxies lie near the KPZ sector, where neither effect dominates. The agreement is good for most of the reported cities. The main outliers are those with very small $\beta$ and $1/z$, which in Fig.~\ref{fig:phase}(b) also exhibit weak overall growth. These cases are plausibly close to a nearly saturated regime, where both front dynamics and roughening are weak and the local roughness has already relaxed near $1/2$. Together with the exponent cloud shown in Fig.~\ref{fig:phase}(a), this ordering supports the two-field Eden scenario proposed here, while also making clear that $p_{\rm emp}$ and $q_{\rm emp}$ should be viewed as effective observables rather than microscopic probabilities.

Our mechanism also clarifies why no direct identification with standard quenched KPZ depinning is appropriate. In the present model the underlying cluster admits overhangs and the near-threshold geometry is isotropic. The relevant critical object is therefore the ordinary-percolation backbone controlling the projected front, not a directed single-valued elastic line. This distinction is essential for the urban problem, where branch growth and mergers are empirically ubiquitous.

\emph{Conclusion}.—We have identified a disorder-crossover mechanism by which universal local roughness and nonuniversal effective dynamics can coexist in projected fronts. In the minimal Eden realization studied here, quenched dilution drives an isotropic morphological transition governed by ordinary percolation, while quenched acceleration modifies both the traversal-time metric and the initial correlated state of the projected front. Under the scaling hypothesis $z_d^{(\rm ct)}\to d_{\min}$, the disorder-to-KPZ crossover obeys $t_{\mathrm{cross}}\sim (\pc-p)^{-\nu d_{\min}}$, while a single-scale critical-height ansatz gives $W(t_{\rm ct})\sim t_{\rm ct}^{1/d_{\min}}$ in the propagation clock and $W(t_{\rm mass})\sim t_{\rm mass}$ in the mass clock. Throughout this regime, $\alpha_{\rm loc}$ stays near $1/2$ because the short-distance sector of the projected front remains Brownian-like even before global saturation.

The significance of this result is broader than the urban application that motivated it. We do not propose a separate universality class for each city; rather, we show how one local-growth mechanism can generate a wide cloud of effective $(\beta,1/z)$ values under finite-time observation. The same logic may apply, with system-specific reinterpretations of dilution and acceleration, to other rough advancing fronts in heterogeneous media whenever the measured object is a projected front and the microscopic disorder separately controls accessibility and traversal times. Urban sprawl is especially valuable because it provides a large-scale empirical test bed, but the underlying mechanism speaks equally to range expansions~\cite{chu2019evolution,horowitz2019bacterial}, colony growth~\cite{huergo2010morphology,swartz2024new}, and deposition problems~\cite{almeida2014universal,orrillo2017morphological} in which long crossover regimes can mimic nonuniversal asymptotic scaling.

The present theory is intentionally minimal and therefore has clear limitations. In particular, urban coalescence is modeled here by a quenched acceleration field, whereas real cities merge with other urban areas that are themselves growing and reshaping the surrounding landscape. Likewise, our model uses strictly local growth rules and uncorrelated quenched disorder. Important next steps are therefore to replace the static acceleration field by interacting secondary growth centers, to study nonlocal growth rules inspired by gravity-type urban models~\cite{ribeiro2023mathematical}, and to determine how spatially correlated disorder or noise modifies the crossover scenario.

\begin{acknowledgments}
\emph{Acknowledgments} -- The authors would like to thank Professor Paolo de Los Rios and Professor Florent Krzakala for their enriching discussions regarding this project. %G.M. acknowledges support from the Swiss National Science Foundation (SNSF) Weave/Lead Agency funding scheme (grant No. 213995).
\end{acknowledgments}

\emph{Author contribution} -- M.H. and M.T. designed the model, analysed and interpreted the simulations. M.H wrote the original draft. G.M. supervised the research. M.H., M.T and G.M. interpreted and discussed the results, and reviewed and edited the manuscript.

The authors declare they have no conflicts of interest.

\emph{Data availability} -- The dataset used in this work, the World settlement footprint (WSF) evolution, is freely accessible from the German Aerospace Center (DLR), Earth Observation
Center \cite{wsf_evolution}.

\bibliography{ref}

\clearpage

\onecolumngrid
\setcounter{section}{0}
\setcounter{figure}{0}
\setcounter{table}{0}
\setcounter{equation}{0}
\renewcommand{\thefigure}{S\arabic{figure}}
\renewcommand{\thetable}{S\arabic{table}}
\renewcommand{\theequation}{S\arabic{equation}}
\section*{Supplemental Material}
\subsection{Supplementary Figures}

\begin{figure}[h!]
    \centering    \includegraphics[width=.7\linewidth]
    {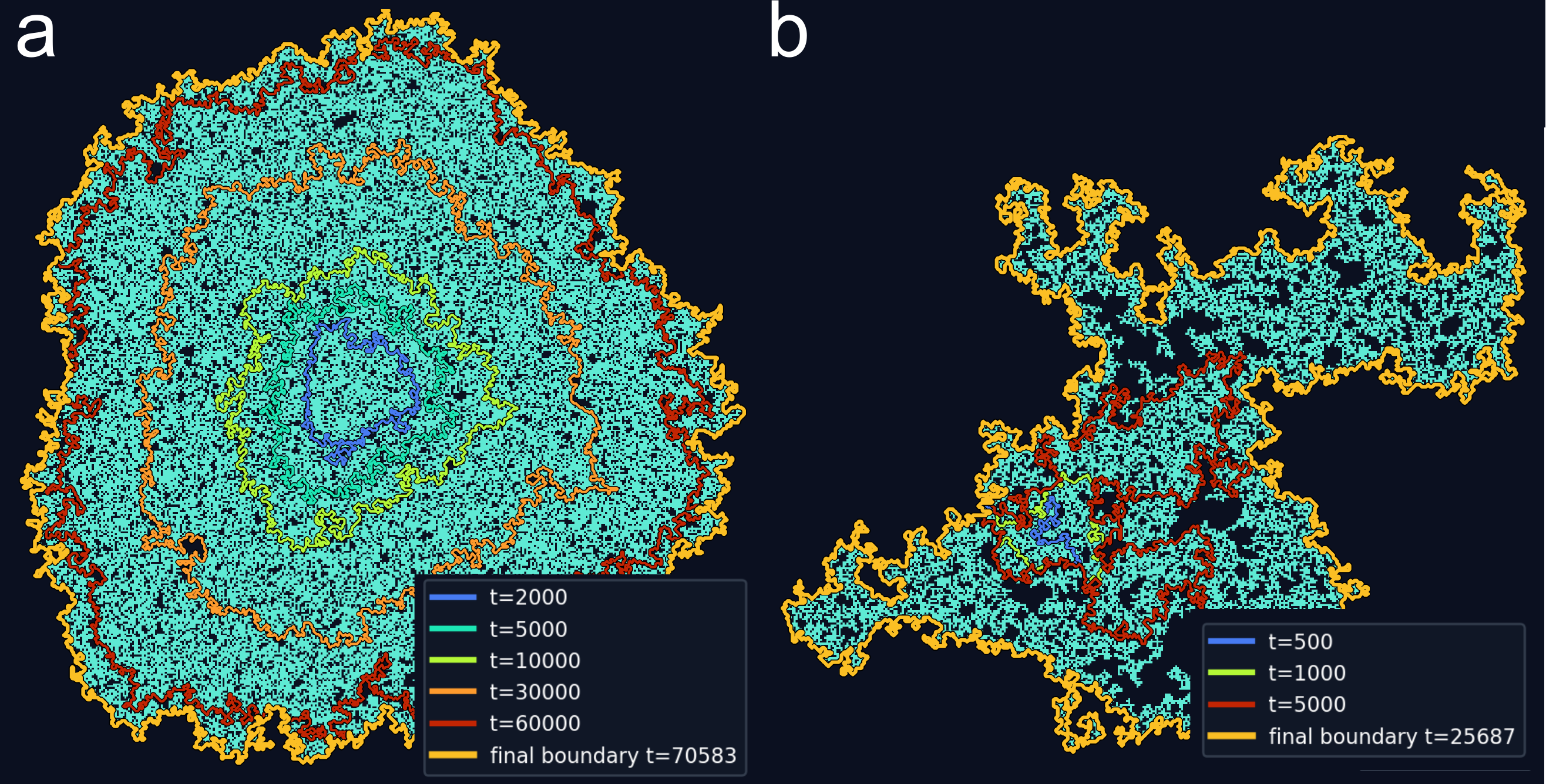}    \caption{Representative realizations of the diluted--accelerated Eden model in radial geometry. Occupied sites are shown together with the cluster boundary at successive times. (a) $p=0.1$, $\rho_a=(1-p)q=0.27$. (b) $p=0.4$, $\rho_a=0.48$. The increase in $p$ and $\rho_a$ leads to a more channelized morphology.}
    \label{fig_si:illu_radial}
\end{figure}

\begin{figure}[h!]
    \centering    \includegraphics[width=.6\linewidth]{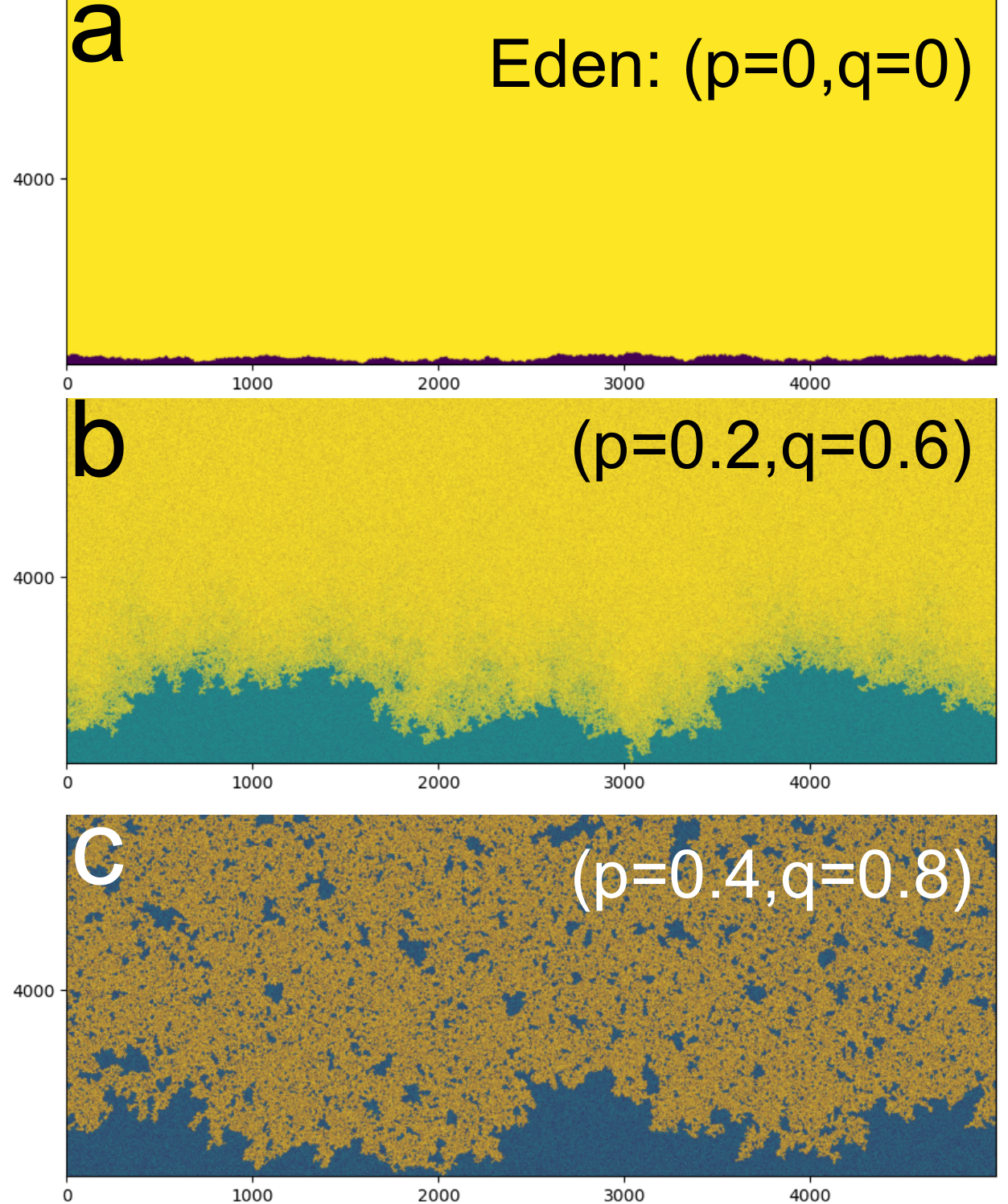} 
    \caption{Final clusters of the diluted--accelerated Eden model in strip geometry at $t=10^8$ on a $5000\times5000$ lattice (yellow site are occupied). (a) $p=0$, $\rho_a=0$. (b) $p=0.2$, $\rho_a=0.48$. (c) $p=0.4$, $\rho_a=0.48$. Panels (b) and (c) have the same acceleration density but different dilution, illustrating the distinct effects of accelerated growth and accessibility disorder on the cluster morphology.}

    \label{fig_si:illu_strip}
\end{figure}

\begin{figure}[h!]
    \centering    \includegraphics[width=.8\linewidth]{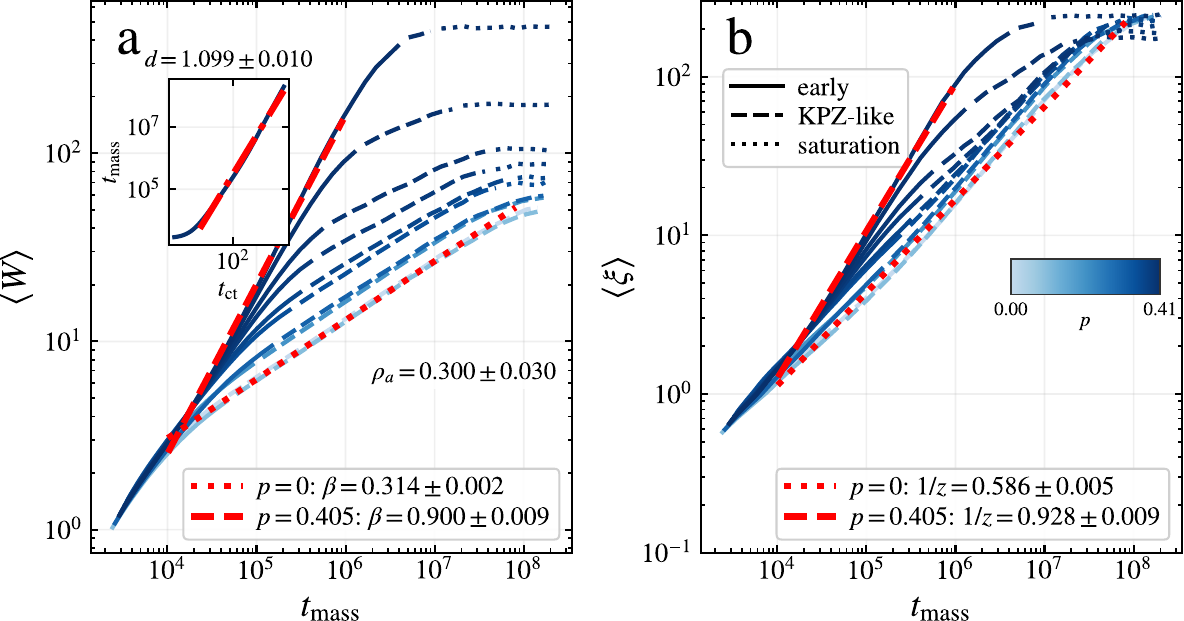}    
    \caption{Global-width and correlation-length scaling with acceleration fields of density $\rho_a\approx0.3   
    $. Data are averaged ($\langle \cdot \rangle$) over $500$ realizations for a strip of width $L=5000$. (a) Projected global width $W(t_{\rm mass})$ for representative values of $p<p_c$. (b) Correlation length $\xi(t_{\rm mass})$ for the same values of $p$. }
    \label{fig_si:width_acc}
\end{figure}
%(c,d) Collapse of the width data and the correlation-length data Eqs.~\ref{eq:Wscaling} and \ref{eq:xiscaling}.
\begin{figure}[h!]
    \centering
    \includegraphics[width=1\linewidth]{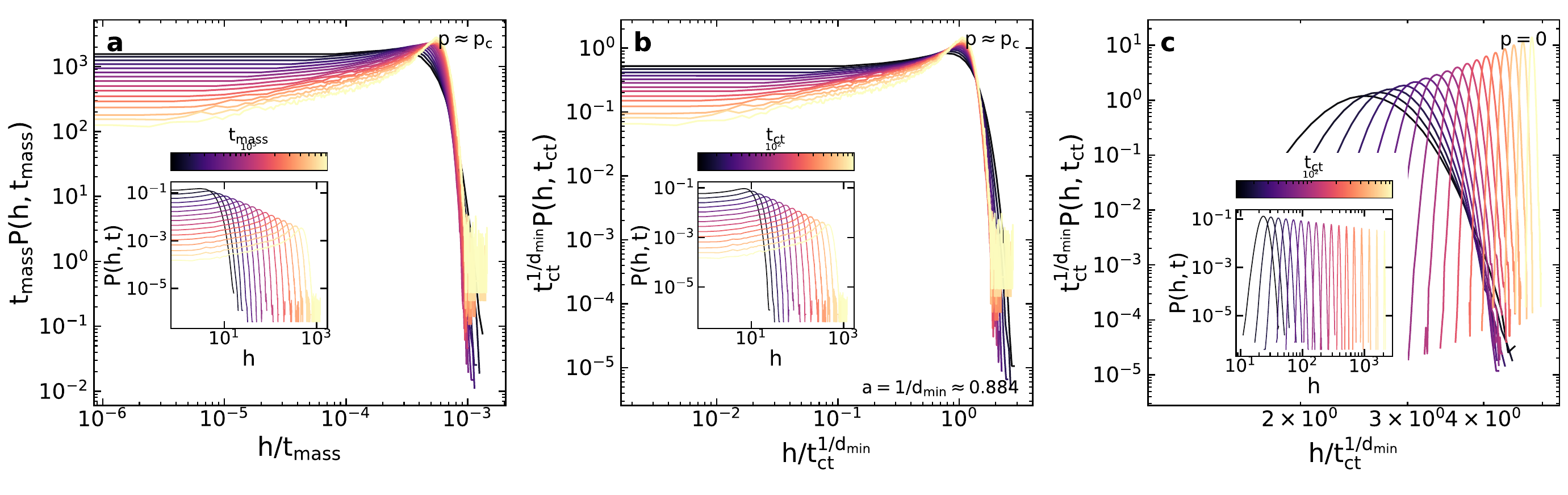}
    \caption{Single-scale collapse of the projected-height distribution $P(h,t)$ for $q=0$. (a) Near $p_c$, the distributions collapse well in the mass clock $t_{\rm mass}$ under the scaling form $P(h,t_{\rm mass})=t_{\rm mass}^{-1}f(h/t_{\rm mass})$. (b) The same near-critical data collapse well in the propagation clock $t_{\rm ct}$ under $P(h,t_{\rm ct})=t_{\rm ct}^{-1/d_{\min}}f(h/t_{\rm ct}^{1/d_{\min}})$, Eq.\ref{eq:single}. Insets show the corresponding raw distributions. (c) Applying the $t_{\rm ct}$ collapse to the clean case $p=0$ does not produce an equally good collapse, showing that the single-scale percolative form is characteristic of the near-critical disordered regime rather than of clean Eden growth.}
    \label{fig_si:pdf_h_collapse}
\end{figure}

\begin{figure}
    \centering
    \includegraphics[width=1\linewidth]{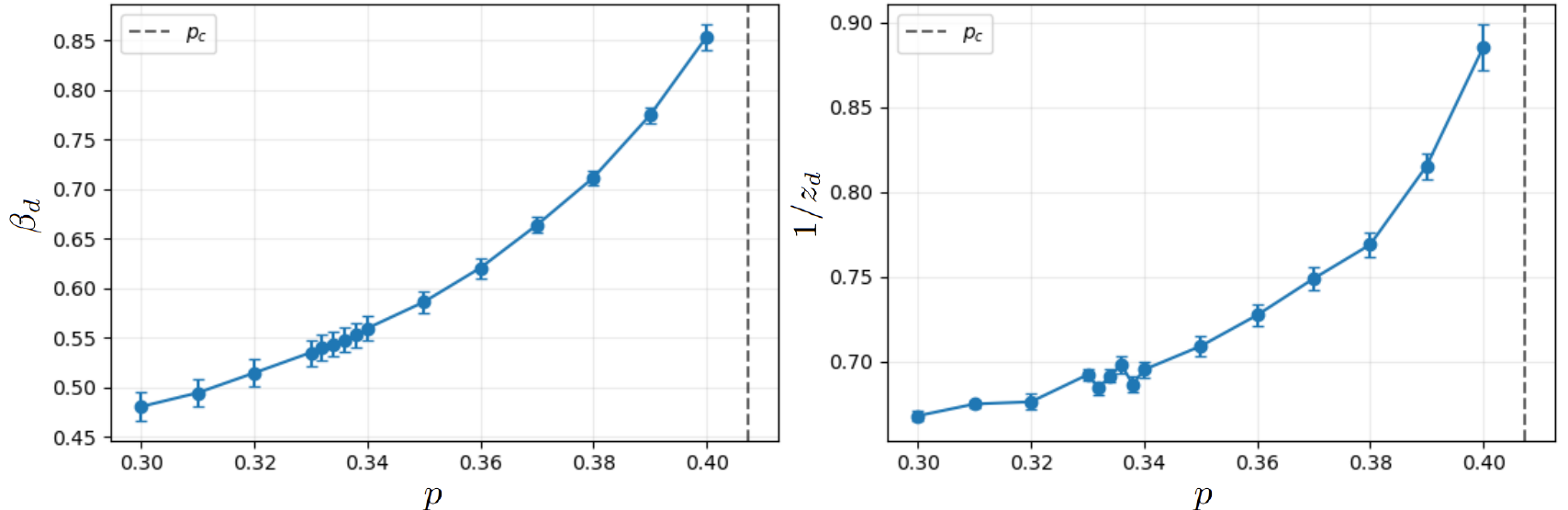}
\caption{Effective exponents in the dilution-only case $q=0$ as a function of $p$, using the propagation clock $t_{\rm ct}$. (a) Transient growth exponent $\beta_d$ extracted in the disorder-dominated window after the initial Poissonian regime and before crossover to KPZ growth or saturation cutoff. (b) Corresponding effective exponent $1/z_d$ from the correlation-length growth over the same window. Both exponents vary continuously with $p$, consistent with a crossover interpretation of the disordered regime.}    \label{fig_si:beta_invz_vs_p}
\end{figure}

\begin{figure}[h!]
    \centering
    \includegraphics[width=.7\linewidth]{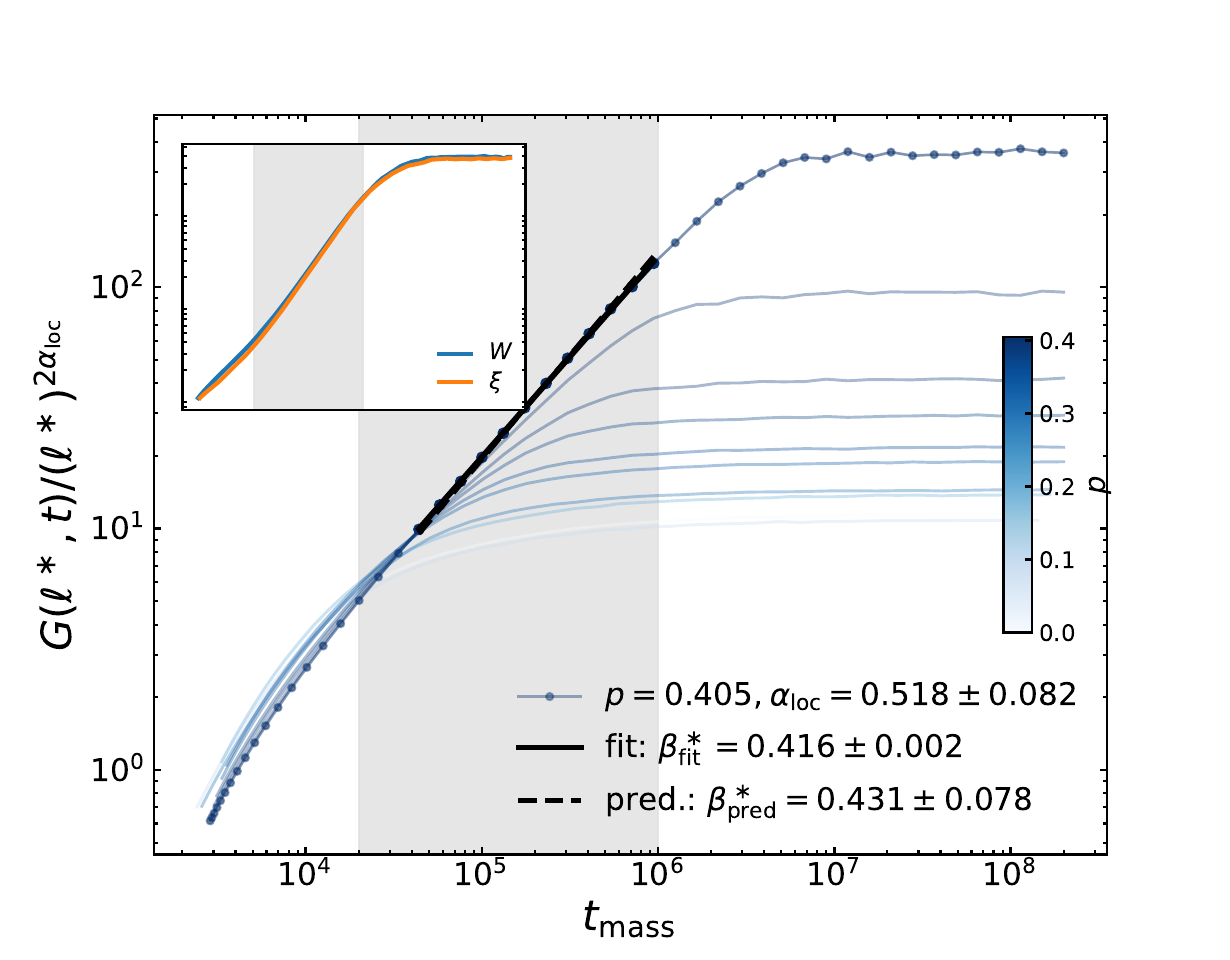}   
\caption{ Compensated local correlation in the dilution-only case $\rho_a\approx 0.3$ for several values of $p<p_c$, colored by $p$. The gray band marks the fitting window; the emphasized curve corresponds to the largest $p$. Solid and dashed black lines show the fitted and predicted local-growth laws, respectively. Inset: normalized $W(t)$ and $\xi(t)$ for the same largest-$p$ dataset.}
    \label{fig_si:beta_star_acc}
\end{figure}

% \begin{figure}[h!]
%     \centering
%     \includegraphics[width=1\linewidth]{img/simu_pure_dilution_beta_inv_z.png}   
% \caption{Effective growth and dynamic exponent in function of the dilution probability $p$ for $q=0$ in the disordered regime.}
%     \label{fig_si:simu_beta_inv_z}
% \end{figure}

\subsection{Model implementation}

We simulate Eden growth on a two-dimensional square lattice in strip geometry, with periodic boundary conditions along the lateral direction $x$ and growth along $y$. The initial condition is a fully occupied substrate at $y=0$. At any time, the growth front is represented by the set of active perimeter sites, namely empty nearest neighbors of the occupied cluster that are accessible to growth. A stochastic Eden event consists in selecting one active perimeter site uniformly at random and occupying it. Overhangs are allowed in the underlying cluster. The projected front used in the main text is extracted from the cluster through
\begin{equation}
h(x,t)=\max\{y:(x,y)\ \text{occupied at time }t\}.
\end{equation}

Two quenched fields are introduced. First, each site with $y>0$ is blocked with probability $p$ and can never be occupied. Second, among the nonblocked sites, each site is declared accelerated with probability $q$. When an accelerated site is first touched by the growing cluster, it is occupied instantaneously; if this newly occupied site touches further accelerated sites, the corresponding connected accelerated region is absorbed recursively at the same clock value. In this way, dilution changes accessibility, whereas acceleration changes the traversal-time structure on the accessible open cluster. For $q=0$ the model reduces to Eden growth in a diluted medium.

The simulations are performed in strip geometry with lateral size
\begin{equation}
L=5000,
\end{equation}
and for each parameter pair $(p,q)$ we generate
\begin{equation}
N_{\rm real}=500
\end{equation}
independent realizations. Each realization is evolved up to
\begin{equation}
T_{\rm main}=10^8
\end{equation}
stochastic perimeter events. In the implementation, a site classification cache is used so that each lattice site is assigned once and for all as blocked, normal open, or accelerated open. A large vertical cutoff is imposed only as a numerical safeguard; realizations hitting this cutoff are discarded from the averages.

We record observables using three clocks. The first is the stochastic event count $t_{\rm evt}$, which counts only perimeter-selection events. Instantaneous accelerated cascades do not consume event time. The second is the continuous-time clock
\begin{equation}
t_{\rm ct}(n)=\sum_{m=1}^{n}\frac{1}{N_{\rm perim}(m)},
\end{equation}
where $N_{\rm perim}(m)$ is the number of active perimeter sites just before the $m$th stochastic event. This is the natural physical time of the asynchronous Eden process. The third is the mass clock
\begin{equation}
t_{\rm mass}\equiv M,
\end{equation}
defined as the cumulative number of occupied sites added above the substrate. In the absence of acceleration, $t_{\rm mass}=t_{\rm evt}$ exactly, whereas for $q>0$ the two clocks differ because a single stochastic event can trigger the instantaneous absorption of a whole accelerated cluster. In the main text we use $t_{\rm mass}$ when comparing with the urban measurements, since it is the closest analogue of the population clock used in Ref.~\cite{marquis2025universal}, and we use $t_{\rm ct}$ to expose the propagation properties of the disordered regime.

The global width is computed online during the simulation from the projected profile,
\begin{equation}
W(t)=\sqrt{\langle h^2\rangle-\langle h\rangle^2},
\end{equation}
where the average is taken over the strip. The height-height correlation function is evaluated from stored front profiles using a Fourier-based estimator,
\begin{equation}
G(r,t)=\left\langle [h(x+r,t)-h(x,t)]^2\right\rangle,
\end{equation}
from which we extract the dynamic correlation length through the fixed-fraction condition
\begin{equation}
G(\xi,t)=c\,G_\infty(t),
\qquad
G_\infty(t)=2W^2(t),
\end{equation}
with $c=0.30$ in the simulations reported here. The local roughness exponent is obtained from the short-distance scaling
\begin{equation}
G(r,t)\sim r^{2\alpha_{\rm loc}}.
\end{equation}

The global width is sampled on a logarithmic grid of event times, while correlation functions and correlation lengths are computed from $50$ stored snapshots distributed logarithmically up to $T_{\rm main}$. All reported averages are taken over the completed realizations for a given $(p,q)$ pair.

\subsection{Empirical construction of the effective proxies $p_{\rm emp}$ and $q_{\rm emp}$}
\label{sec:SM_proxies}

\subsection{Urban-settlement data and final-city mask}

For each city we use a two-dimensional raster in which each pixel stores the first year at which that location became occupied by settlement. Pixels with value $0$ are never occupied within the observation window, while positive values indicate the year of first occupation. In practice, the occupied mask at year $Y$ is therefore
\begin{equation}
\mathcal O_Y=\{x:\; t_{\rm occ}(x)\neq 0,\; t_{\rm occ}(x)\le Y\}.
\end{equation}
The final urban object is defined from the mask at the last available year, here $Y=2015$. From $\mathcal O_{2015}$ we extract the largest connected component (LCC), denoted $\mathcal L_{2015}$, using four-neighbor connectivity throughout. All empirical proxies reported below are measured inside this final connected urban component or relative to its tracked ancestry.

\subsection{Tracked ancestry of the final LCC}

To reconstruct the growth history of the final city, we do not recompute the largest connected component independently at each year, since the identity of the instantaneous LCC may switch between disconnected clusters. Instead, we track the ancestry of $\mathcal L_{2015}$ backward in time. Let $\mathcal A_Y$ denote the ancestor of the final LCC at year $Y$. Starting from
\begin{equation}
\mathcal A_{2015}=\mathcal L_{2015},
\end{equation}
we move backward year by year. At each year $Y<2015$, we restrict the occupied mask to the final city,
\begin{equation}
\widetilde{\mathcal O}_Y=\mathcal O_Y\cap \mathcal L_{2015},
\end{equation}
decompose $\widetilde{\mathcal O}_Y$ into connected components, and define $\mathcal A_Y$ as the component with largest overlap with $\mathcal A_{Y+1}$. This yields a continuous lineage for the final urban core and provides the reference object used below to define both internal cavities and future coalescence reservoirs.

\subsection{Effective dilution proxy $p_{\rm emp}$}

The empirical dilution proxy is designed to quantify the amount of empty-space structure generated during growth inside the final urban object. We first define the hole set of a binary mask $\mathcal M$ as
\begin{equation}
\mathcal H(\mathcal M)=\mathrm{Fill}(\mathcal M)\setminus \mathcal M,
\end{equation}
where $\mathrm{Fill}(\mathcal M)$ denotes the mask after filling all enclosed cavities. Thus $\mathcal H(\mathcal M)$ contains empty clusters that are topologically enclosed by $\mathcal M$ and not connected to the exterior.

We then compare the holes of the final city, $\mathcal H(\mathcal L_{2015})$, with the holes already present in an early tracked ancestor, typically $\mathcal H(\mathcal A_{1985})$. The emergent-hole set is defined as
\begin{equation}
\mathcal H_{\rm emerg}
=
\mathcal H(\mathcal L_{2015})\setminus \mathcal H(\mathcal A_{1985}),
\end{equation}
namely the part of the final internal void structure that was not already enclosed in the early ancestor. Our empirical dilution proxy is the corresponding area fraction,
\begin{equation}
p_{\rm emp}
\equiv
\frac{|\mathcal H_{\rm emerg}|}{|\mathcal L_{2015}|}.
\label{eq:SM_pemp}
\end{equation}
This quantity is an effective geometric proxy rather than a microscopic occupation probability: it measures the fraction of the final city occupied by empty cavities generated during the growth process, and therefore captures the remnant of the barrier landscape that survives as internal void structure.

\subsection{Effective coalescence proxy $q_{\rm emp}$}

The empirical coalescence proxy is designed to quantify the density of already-built clusters that are not yet connected to the tracked urban core but will later be absorbed by it. For a given early year $Y$, we define the absorbable set
\begin{equation}
\mathcal B_Y
=
\mathcal O_Y \cap \mathcal L_{2015}\cap \overline{\mathcal A_Y},
\end{equation}
where $\overline{\mathcal A_Y}$ denotes the complement of the tracked ancestor inside the raster. Thus $\mathcal B_Y$ contains pixels that are already urbanized at year $Y$, belong to the final 2015 city, but are not yet part of the tracked main lineage at that year. These are precisely the pre-existing urban patches that can later merge with the main component by coalescence.

For each city we convert this set into a density by normalizing to the area of the final LCC,
\begin{equation}
\rho_{\rm abs}(Y)=\frac{|\mathcal B_Y|}{|\mathcal L_{2015}|}.
\end{equation}
In the main text we use the mean over the early years considered,
\begin{equation}
q_{\rm emp}
\equiv
\overline{\rho}_{\rm abs}
=
\frac{1}{N_Y}\sum_Y \rho_{\rm abs}(Y),
\label{eq:SM_qemp}
\end{equation}
although maximum and first-year versions can be defined analogously. As for $p_{\rm emp}$, this quantity is an effective geometric proxy rather than a microscopic site probability. It measures the reservoir of disconnected urbanized patches available for later absorption by the tracked lineage and therefore captures the coalescence tendency of the observed city.

\subsection{Interpretation}

The proxies $p_{\rm emp}$ and $q_{\rm emp}$ are not meant to coincide numerically with the microscopic disorder probabilities $p$ and $q$ of the diluted--accelerated Eden model. Rather, they provide two empirical control directions that mirror the two ingredients of the model: $p_{\rm emp}$ quantifies empty-space disorder generated inside the final city, while $q_{\rm emp}$ quantifies the occupied-space reservoir of future coalescence. Their role is therefore to organize the empirical exponent cloud into barrier-dominated, coalescence-dominated, and mixed sectors, and not to furnish a one-to-one inversion from data to microscopic model parameters.

\end{document}